\documentclass
[a4paper,aps,pra,twocolumn,amsmath,amssymb]{revtex4}

\usepackage{epsfig}
\usepackage{amsmath}
\usepackage{graphicx}
\usepackage{bm}

\begin{document}

\title{Demonstration of an ultracold micro-optomechanical oscillator in a cryogenic cavity}\thanks{This work was published in Nature Phys. \textbf{5}, 485 (2009).}
\author{Simon Gr\"oblacher,$^{1,2}$ Jared B. Hertzberg,$^{3,4}$ Michael R. Vanner,$^{1,2}$ Garrett D. Cole,$^{1,5}$ Sylvain Gigan,$^{6}$ K. C. Schwab,$^{3,}$\footnotemark\footnotetext{Permanent address:\ Department of Applied Physics, Caltech, Pasadena, CA 91125, USA} and Markus Aspelmeyer$^{1,}$}\email{markus.aspelmeyer@quantum.at}
\affiliation{
$^1$ Institute for Quantum Optics and Quantum Information (IQOQI), Austrian Academy of Sciences, Boltzmanngasse 3, A-1090 Vienna, Austria\\
$^2$ Faculty of Physics, University of Vienna, Boltzmanngasse 5, A-1090 Vienna, Austria\\
$^3$ Department of Physics, Cornell University, Ithaca, NY 14853, USA\\
$^4$ Department of Physics, University of Maryland, College Park, MD 20742, USA\\
$^5$ The Center for Micro- and Nanostructures (ZMNS), Vienna University of Technology, Floragasse 7, A-1040 Vienna, Austria\\
$^6$ Laboratoire Photon et Mati\`{e}re, Ecole Superieure de Physique et de Chimie Industrielles, CNRS-UPRA0005, 10 rue Vauquelin, 75005 Paris, France}

\begin{abstract}
Preparing and manipulating quantum states of mechanical resonators is a highly interdisciplinary undertaking that now receives enormous interest for its far-reaching potential in fundamental and applied science~\cite{Schwab2005,Aspelmeyer2008}. Up to now, only nanoscale mechanical devices achieved operation close to the quantum regime~\cite{LaHaye2004,Naik2006}. We report a new micro-optomechanical resonator that is laser cooled to a level of 30 thermal quanta. This is equivalent to the best nanomechanical devices, however, with a mass more than four orders of magnitude larger (43~ng versus 1~pg) and at more than two orders of magnitude higher environment temperature (5~K versus 30~mK). Despite the large laser-added cooling factor of 4,000 and the cryogenic environment, our cooling performance is not limited by residual absorption effects. These results pave the way for the preparation of 100-$\mu$m scale objects in the quantum regime. Possible applications range from quantum-limited
optomechanical sensing devices to macroscopic tests of quantum physics~\cite{Marshall2003,Vitali2007}.
\end{abstract}

\maketitle

Recently, the combination of high-finesse optical cavities with mechanical resonators has opened up new possibilities for preparing and detecting mechanical systems close to\textemdash and even in\textemdash the quantum regime by using well-established methods of quantum optics. Most prominently, the mechanism of efficient laser cooling has been demonstrated~\cite{Metzger2004,Gigan2006,Arcizet2006b,Schliesser2008,Corbitt2007,Thompson2008,Teufel2008} and has been shown to be capable, in principle, of reaching the quantum ground state~\cite{Wilson-Rae2007,Marquardt2007,Genes2008}. A particularly intriguing feature of this approach is that it can be applied to mechanical objects of almost arbitrary size, from the nanoscale in microwave strip-line cavities~\cite{Teufel2008} up to the centimetre scale in gravitational-wave interferometers~\cite{Corbitt2007}. In addition, whereas quantum-limited readout is still a challenging development step for non-optical schemes~\cite{LaHaye2004,Regal2008,Poggio2008}, optical readout techniques at the quantum limit are readily available~\cite{Arcizet2006a}.\\

Approaching and eventually entering the quantum regime of mechanical resonators through optomechanical interactions
essentially requires the following three conditions to be fulfilled: (1) sideband-resolved operation; that is, the cavity amplitude decay rate $\kappa$ has to be small with respect to the mechanical frequency $\omega_m$; (2) both ultralow noise and low absorption of the optical cavity field (phase noise at the mechanical frequency can act as a
finite-temperature thermal reservoir and absorption can increase the mode temperature and even diminish the cavity performance in the case of superconducting cavities); and (3) sufficiently small coupling of the mechanical resonator to the thermal environment; that is, low environment temperature $T$ and large mechanical quality factor $Q$ (the thermal coupling rate is given by $k_BT/\hbar Q$, where $k_B$ is the Boltzmann constant and $\hbar$ is the reduced Planck constant). So far, no experiment has demonstrated all three requirements simultaneously. Criterion (1) has been achieved~\cite{Schliesser2008,Teufel2008,Cole2008}; however, the performance was limited in one case by laser phase noise~\cite{Schliesser2008} and in the other cases by absorption in the cavity~\cite{Teufel2008,Cole2008}. Other, independent, experiments have implemented only criterion (2)~\cite{Corbitt2007,Thompson2008,Arcizet2006a,Groeblacher2008}. Finally, criterion (3) has been realized in several cryogenic experiments~\cite{Naik2006,Teufel2008,Groeblacher2008,Poggio2007}, however not in combination with both (1) and (2).\\

\begin{figure*}[htbp]
\centerline{\includegraphics[width=0.96\textwidth]{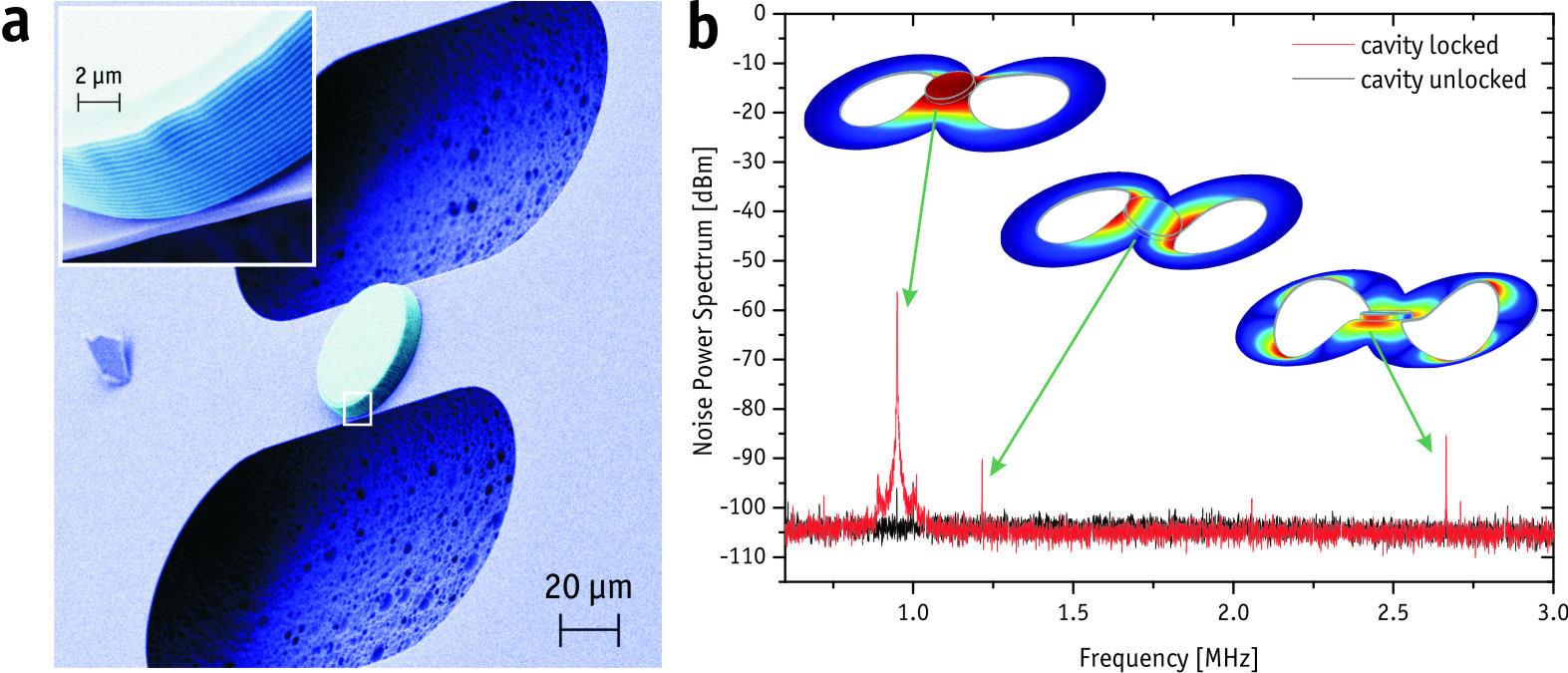}} \caption{\textbf{High-quality micro-optomechanical resonator. a}, Scanning electron micrograph of the basic mechanical system, which is formed by a doubly clamped Si$_3$N$_4$ beam. A circular, high-reflectivity Bragg mirror is used as the end mirror of a Fabry-P\'{e}rot cavity. The Bragg mirror is made of low-absorption, alternating dielectric stacks of Ta$_2$O$_5$/SiO$_2$. The magnified section in the inset shows the stacking sequence. \textbf{b}, Micromechanical displacement spectra shown as noise power spectra of the readout-beam phase quadrature for a locked and an unlocked cavity. The fundamental mode at $\omega_m=2\pi\times 945$~kHz and all higher mechanical modes are identified by finite element simulation. For the cases that involve large Bragg mirror displacements, we provide the simulated mode profile.} \label{fig1}
\end{figure*}

We have designed a novel micro-optomechanical device that enables us to meet all requirements at the same time. Specifically, we have fabricated a Si$_3$N$_4$ micromechanical resonator that carries a high-reflectivity, ultralow-loss Bragg mirror (Fig.\ 1a), which serves as the end mirror of a Fabry-P\'{e}rot cavity. We designed the
system to exhibit a fundamental mechanical mode at relatively high frequency (of the order of 1~MHz; Fig.\ 1b) such that sideband-resolved operation (criterion (1)) can be achieved already with a medium-finesse cavity. Criterion (2) can first be fulfilled because our solid-state pump laser used for optical cooling exhibits low phase noise (laser linewidth below 1~kHz). Second, absorption in the Bragg mirror is sufficiently low to prevent residual heating in
the mechanical structure. Absorption levels as low as $10^{-6}$ have been reported for similar Bragg mirrors~\cite{Rempe1992} and recent measurements suggest even lower values of $4\times 10^{-7}$ for the specific coatings used in this experiment (R. Lalezari, private communication). In addition, although absorption in Si$_3$N$_4$ is comparable to silicon, the transmission mismatch of the two cavity mirrors ($\sim$10:1) and the resulting low transmission through the Bragg mirror prevents residual heating of the resonator as has been observed
for cryogenically cooled silicon cantilevers~\cite{Bleszynski-Jayich2008}. Finally, criterion (3) requires low temperature and high mechanical quality. The mechanical properties of our design are dominated by the Si$_3$N$_4$,
which is known to exhibit superior performance in particular at low temperatures, where $Q$-factors beyond $10^6$ have been observed at millikelvin temperatures~\cite{Zwickl2008}.\\

We operate our device, a $100~\mu\mathrm{m}\times50~\mu\mathrm{m}\times 1~\mu\mathrm{m}$ microresonator, in a cryogenic $^4$He environment at $10^{-7}$~mbar and in direct contact with the cryostat cold finger. To measure the mechanical displacement, the frequency of a 7~$\mu$W continuous-wave Nd:YAG laser is locked close to resonance of the cryogenic Fabry-P\'{e}rot
cavity (length $L\approx25$~mm), which consists of a fixed macroscopic mirror and the moving micromechanical mirror. The optical cavity of finesse $F\approx 3,900$ achieves moderate sideband resolution ($\kappa\approx 0.8\omega_m$), which in principle would allow cooling to a final occupation number $\langle n\rangle_{min}=(\frac{\kappa^2}{4\omega_m^2})\approx 0.16$, that is, well into the quantum ground state~\cite{Wilson-Rae2007,Marquardt2007}. The experimentally achievable temperature is obtained as the equilibrium state of two competing processes, namely the laser cooling rate and the coupling rate to the thermal (cryogenic) environment. In essence, laser cooling is driven (in the ideal resolved-sideband limit and at detuning $\Delta=\omega_m$) at a rate $\Gamma\approx G^2/(2\kappa)$ ($G$ is the effective optomechanical coupling rate, as defined in ref.~\cite{Genes2008}), whereas mechanical relaxation to the thermal environment at temperature $T$ takes place at a rate ($k_BT/\hbar Q$). The final achievable mechanical occupation number is therefore, to first order, given by $n_f\approx(1/\Gamma)\times(k_BT/\hbar Q)$. A more accurate derivation taking into account effects of non-ideal sideband resolution can be found, for example, in refs~\cite{Wilson-Rae2007,Marquardt2007,Genes2008,Wilson-Rae2008}. Our experimental parameters limit the minimum achievable mode temperature to approximately 1~mK ($n_f\approx 30$). The fact that we can observe this value in the
experiment (see below) shows that other residual heating effects are negligible. The micromechanical flexural motion modulates the cavity-field phase quadrature, which is measured by optical homodyning. For $Q\gg 1$ its noise power spectrum (NPS) is a direct measure of the mechanical position spectrum $S_q(\omega)$, as described in ref.~\cite{Genes2008}. We observe a minimum noise floor of $2.6\times 10^{-17}$~m$\,$Hz$^{-0.5}$, which is a factor of 4 above the achievable quantum (shot-noise) limit, when taking into account the finite cavity linewidth, the cavity losses and the non-perfect mode-matching, and due to the residual amplitude noise of the pump laser at the sideband frequency of
our mechanical mode. We observe the fundamental mechanical mode at $\omega_m=945$~kHz with an effective mass $m_{eff}=(43\pm 2)$~ng and a quality factor $Q\approx 30,000$ at 5.3~K ($Q\approx 5,000$ at 300~K). These values are consistent with independent estimates based on finite-element method simulations yielding $\omega_m=945$~kHz and
$m_{eff}=(53\pm 5)$~ng (see the section Effective mass).\\

Optomechanical laser cooling requires driving of the cavity with a red-detuned (that is, off-resonant), optical field~\cite{Vitali2007,Metzger2004,Gigan2006,Arcizet2006b,Schliesser2008,Corbitt2007,Thompson2008,Teufel2008}. We achieve this by coupling a second laser beam\textemdash detuned by $\Delta$ in frequency but orthogonal in polarization\textemdash into the same spatial cavity mode (Fig.\ 2a). Birefringence of the cavity material leads to both an optical path length difference for the two cavity modes (resulting in an 800~kHz frequency difference of the cavity peak positions) and a polarization rotation of the outgoing fields. We compensate both effects by an offset in $\Delta$ and by extra linear optical phase retarders, respectively. A change in detuning $\Delta$ modifies the mechanical rigidity and results in both an optical spring effect ($\omega_{eff}(\Delta)$) and damping ($\gamma_{eff}(\Delta)$), which is directly extracted by fitting the NPS using the expressions from ref.~\cite{Genes2008}. Figure 2b shows the predicted behaviour for several powers of the red-detuned beam. The low-power curve at 140~$\mu$W is used to determine both the effective mass of the mechanical mode, $m_{eff}$, and the cavity finesse, $F$. For higher powers and detunings closer to cavity resonance, the onset of cavity instability prevents a stable lock (see, for example, ref.~\cite{Genes2008}). All experimental data are in agreement with theory and hence in accordance with pure radiation-pressure effects~\cite{Marquardt2007}.\\

\begin{figure*}[htbp]
\centerline{\includegraphics[width=0.96\textwidth]{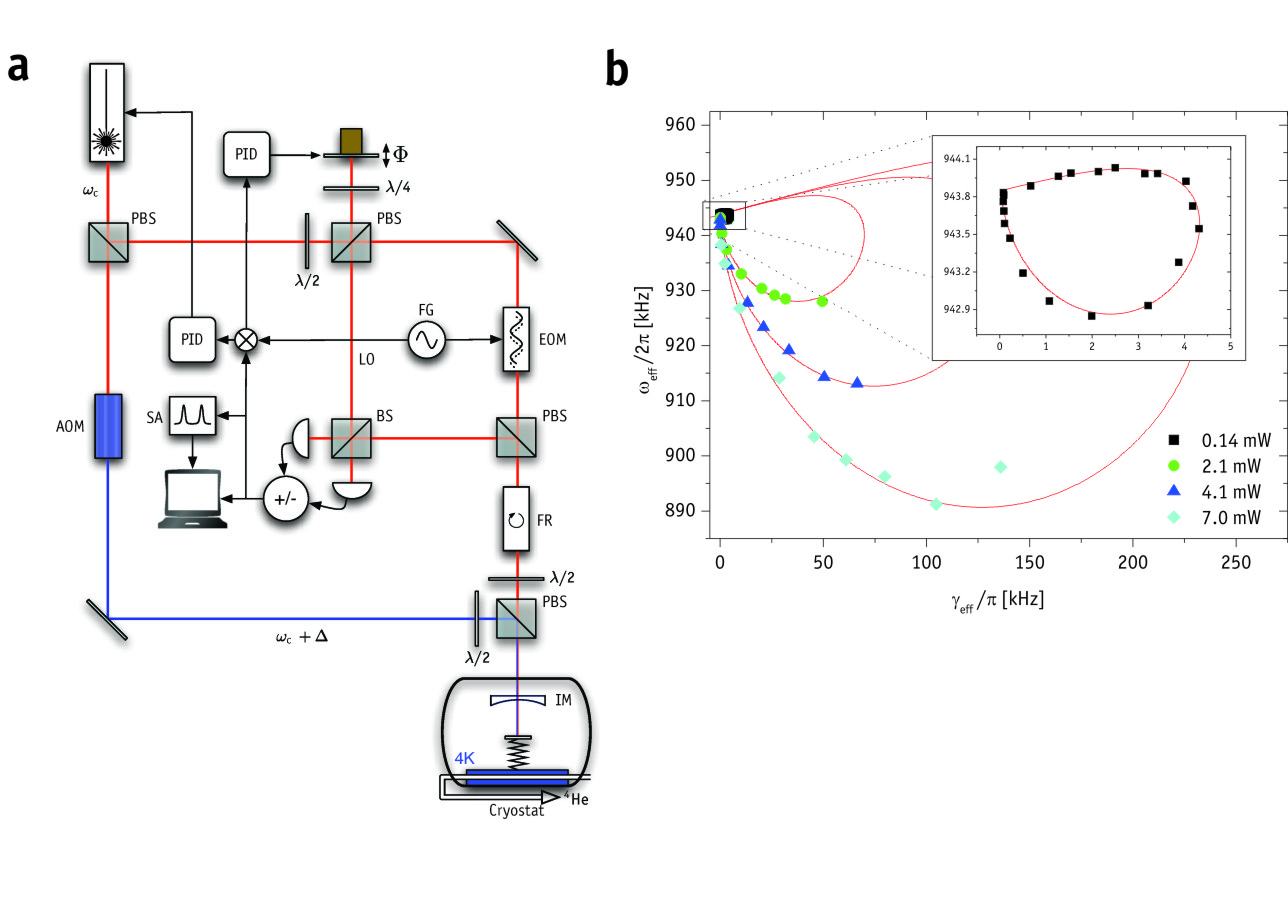}} \caption{\textbf{Experimental set-up and characterization of optomechanical radiation-pressure interaction. a}, The laser is split at a polarizing beamsplitter (PBS) into a weak locking field (red) tuned near cavity resonance $\omega_c$ and the cooling field (blue) tuned off-resonant with an acousto-optical modulator (AOM) to $\omega_c+\Delta\approx\omega_c-\omega_m$. An electro-optical modulator (EOM) in the weak field is used to generate a Pound-Drever-Hall error signal for cavity locking. The beams are recombined on a PBS into the same spatial mode at orthogonal polarization before they enter the cavity comprising an input mirror (IM) and the micro-mechanical mirror. The phase quadrature of the locking beam is measured in a homodyne detection scheme (BS:\ beamsplitter; LO:\ local oscillator; $\Phi$:\ local oscillator phase; SA:\ spectrum analyser). $\Phi$ is stabilized in a separate proportional-integral-derivative controller (PID). A combination of a Faraday rotator (FR) and a half-wave plate ($\lambda$/2) separates the reflected from the original signal. \textbf{b}, The effective frequency $\omega_{eff}$ and damping $\gamma_{eff}$ of the micro-mechanical motion for different detuning and power settings. All power levels follow the theoretical predictions for pure radiation-pressure interaction. The symbols are experimental data, and the solid lines are simulations based on ref.~\cite{Genes2008}. The inset shows the data set taken at $140~\mu$W optical power.} \label{fig2}
\end{figure*}

The effective mode temperature is obtained through the equipartition theorem. For our experimental parameter regime, $Q\gg 1$ and $\langle n\rangle\gg 0.5$, the integrated NPS is also a direct measure of the mean mechanical mode energy and hence, through the equipartition theorem, of its effective temperature through $T_{eff}=(m_{eff}\omega_{eff}^2/k_B)\int^{+\infty}_{-\infty}NPS(\omega)d\omega$. Note that, for the case of strong
optomechanical coupling, normal-mode splitting can occur and has to be taken into account when evaluating the mode temperature~\cite{Dobrindt2008}. In our present case, this effect is negligible because of the large cavity decay rate $\kappa$. The amplitude of the NPS is calibrated by comparing the mechanical NPS with the NPS of a known frequency modulation applied to the laser (see, for example, ref.~\cite{Pinard1999}). For a cold-finger temperature of 5.3~K, we obtain a mode temperature $T=2.3$~K, which is consistent with an expected moderate cooling due to slightly off-resonant locking of the Fabry-P\'{e}rot cavity (by less than 3$\%$ of the cavity intensity linewidth). The locking point is deliberately chosen to be on the cooling side to avoid unwanted parametric mechanical instabilities. The mean thermal occupancy was calculated according to $\langle n\rangle=k_BT_{eff}/\hbar\omega_{eff}$. We note, however, that Bose-Einstein statistics will have a dominant role as one approaches the quantum ground state.\\

Figure 3a shows mechanical noise power spectra with the cooling beam switched off and with maximum cooling beam pump power at 7~mW. For a detuning $\Delta\approx\omega_m$, we demonstrate laser cooling to a mean thermal occupation of $32\pm 4$ quanta, which is more than 2 orders of magnitude lower than previously reported values for optomechanical devices~\cite{Schliesser2008} and is comparable to the lowest reported temperature of 25 quanta for nano-electromechanical systems~\cite{Naik2006} (NEMS). In contrast to previous experiments~\cite{Schliesser2008,Teufel2008}, the achieved cooling performance is not limited by optical absorption
or residual phase noise, but follows exactly the theoretically predicted behaviour (Fig.\ 3b). This agrees with the expected device performance: a fraction of approximately $10^{-6}$ of the intra-cavity power is absorbed by the Bragg mirror ($\sim 13~\mu$W at maximum cooling) and a maximum of $1\%$ of the transmitted power is absorbed by the Si$_3$N$_4$ beam~\cite{Palik1998} ($\sim 14~\mu$W at maximum cooling and taking into account the impedance mismatch of the cavity mirrors). The cryogenic cooling power of the cryostat used is orders of magnitude larger than the maximum heat load expected on the micromechanical structures. The absence of absorption can also be seen from the inferred mode temperature $T_{eff}$, which decreases with the mechanical damping rate $\gamma_{eff}$ in strict accordance with the power law $T_{eff}\propto\gamma_{eff}^{-1}$. This relation follows immediately from the simple expression for the mechanical occupation $n_f$ given above ($n_f\propto T^{-1}$) and from the fact that the laser cooling rate $\Gamma$ is to first approximation equivalent to the effective mechanical damping $\gamma_{eff}$,
at least for all data points of our experiment. Both heating and the onset of normal-mode splitting for strong coupling~\cite{Dobrindt2008} would result in a deviation of this behaviour.\\

\begin{figure*}[htbp]
\centerline{\includegraphics[width=0.96\textwidth]{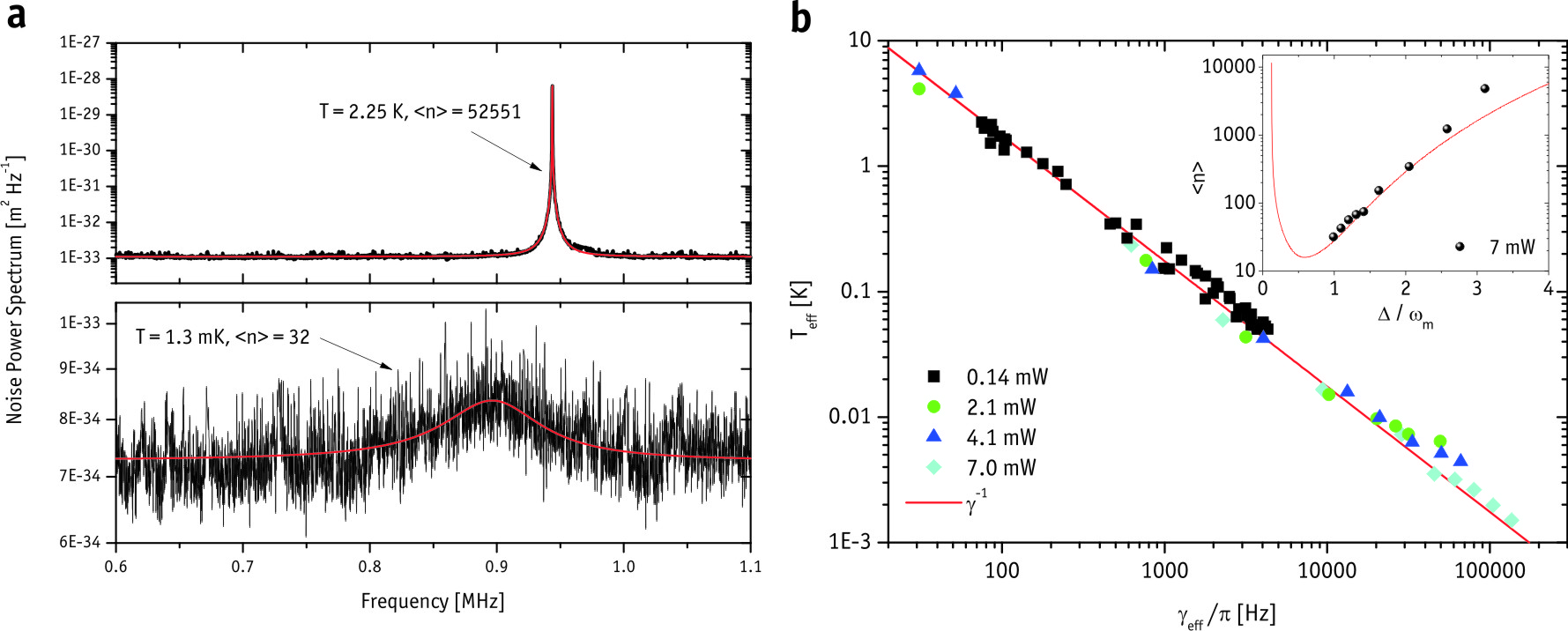}} \caption{\textbf{Optomechanical laser cooling inside a cryogenic cavity. a}, Calibrated noise power spectra for the fundamental mechanical mode at 5.3~K environmental temperature with small cavity cooling (top) and at maximum cooling (bottom). The thermal energy is reduced from $\approx$53,000 quanta at $7~\mu$W laser power to $32\pm 4$ quanta at 7~mW. The vertical axes in both plots are logarithmic. The change in the technical noise floor is due to different locking levels of the local oscillator phase $\Phi$ in the homodyne detection. \textbf{b}, Plot of the calibrated effective temperature $T_{eff}$ versus the observed damping $\gamma_{eff}$ for various power and detuning values of the cooling beam. No deviations from the theoretically expected power-law dependence (red solid line) can be observed. The inset shows the mean thermal occupation $\langle n\rangle$ as a function of detuning for maximal laser power. Cavity instability prevents detunings arbitrarily close to resonance. The red solid curve is a simulation based on ref.~\cite{Genes2008} that uses only experimentally obtained parameters.} \label{fig3}
\end{figure*}

The remaining obstacle that prohibits us from reaching the quantum ground state is the intrinsic phonon coupling to the thermal environment at rate $k_BT/\hbar Q\approx 1.4\times 10^7$~Hz. By reducing the reservoir temperature to that of NEMS experiments (20~mK), this coupling will significantly reduce, not only owing to the lower bath temperature but also because Si$_3$N$_4$ resonators markedly improve in mechanical $Q$ with decreasing temperature. For example, thermal heating rates as low as $3\times 10^3$~Hz have been observed for Si$_3$N$_4$ at 300~mK (ref.~\cite{Zwickl2008}), which would place our effective mode temperature already well into the quantum ground state using otherwise unchanged parameters.\\

In summary, we have demonstrated optical cooling of the fundamental mode of a $100~\mu$m scale mechanical resonator in a cryogenic cavity to a thermal occupation of only $32\pm 4$ quanta. This is comparable to the performance of state-of-theart NEMS devices. In contrast to previous approaches, the large laser cooling rates attained are no longer limited by residual absorption or phase-noise effects. This is achieved by a new micro-optomechanical resonator design with exceptionally low intrinsic optical absorption and both high optical and mechanical quality. This leaves the reduction of the thermal coupling, for example, by further decreasing the environment temperature to those available in conventional $^3$He cryostats, as the only remaining hurdle to prepare the mechanical quantum ground state. Our approach hence establishes a feasible route towards the quantum regime of massive micromechanical systems.\\

\begin{center}
\large\textbf{Methods}\\
\end{center}

\begin{center}
\small\textbf{Micro-mirror fabrication}\\
\end{center}
\normalsize

Our micro-mechanical oscillator is made of 1-$\mu$m-thick low-stress Si$_3$N$_4$ deposited on a Si substrate and coated through ion beam sputtering with a high-reflectivity Bragg mirror. Standard photolithography and plasma
etching is used for forming, in subsequent steps, the mirror pad and the micro-mechanical resonator, which is finally released from the Si substrate in a XeF$_2$ atmosphere. The mirror stack, designed and deposited by ATFilms, comprises 36 alternating layers of Ta$_2$O$_5$ and SiO$_2$ with an overall nominal reflectivity of 99.991$\%$ at 1,064~nm. The measured finesse of 3,900 is consistent with an input coupler reflectivity of 99.91$\%$ and with extra diffraction losses due to a finite size of the cavity beam waist.\\

\begin{center}
\large \textbf{Supplementary Information}\\
\end{center}

\begin{center}
\small\textbf{Effective mass}\\
\end{center}
\normalsize

We have estimated the effective mass of the fundamental mode of our micromechanical structure using both analytic models and FEM analysis. The experimentally observed value of $43\pm 2$~ng agrees to within 10$\%$ with the estimated value of $53\pm 5$~ng.

The total mass of the dielectric Bragg mirror (radius $R\approx 24.5\pm 0.5$~$\mu$m) made of 36 alternating layers of Ta$_2$O$_5$ ($\rho\approx 8,200$~kg/m$^3$, $t=126.4$~nm) and SiO$_2$ ($\rho=2,200$~kg/m$^3$, $t=179.6$~nm) is $45\pm 5$~ng, not taking into account the lateral etch and tapering of the mirror pad. The large error stems from the uncertainty in the exact value of the Ta$_2$O$_5$ density, which can vary between $6,800$ and $8,300$~kg/m$^3$. The mass of the Si$_3$N$_4$ resonator ($\rho=3,000$~kg/m$^3$, approximate dimensions of $100\times 50\times 1~\mu$m$^3$) is approx.\ 11~ng, resulting in a maximum total mass of $56\pm 5$~ng for the full optomechanical device.

The mode mass, i.e.\ the actual mass contributing to the motion of the Si$_3$N$_4$ resonator fundamental mode, is approx.\ 74$\%$ of the total mass of the Si$_3$N$_4$ resonator (see any standard literature on elasticity theory, for example~\cite{Harrington1994}). This would result in a total mode mass of the optomechanical resonator (Si$_3$N$_4$ beam plus micromirror) of approx.\ $53\pm 5$~ng. However, because of the flat-top mode shape of our actual device (see the FEM simulation shown in Figure 4), this value is only a conservative lower bound. A more realistic value that takes into account the actual mode shape can be obtained directly from FEM simulation and is approx.\ $56\pm 5$~ng (see below).

Finally, to calculate the effective mass one has to take into account the mode overlap between the mechanical resonator mode and the mode of the optical probe beam (for a detailed analysis on the calculation of the effective mass see for example~\cite{Pinard1999}). Based on the experimentally obtained optical finesse, which is limited by intensity losses due to a finite mirror size, we can provide an upper bound on the cavity beam waist at the micromirror position of $8\pm 2$~$\mu$m. If we assume a mechanical mode shape of an ideal doubly-clamped beam of dimensions $100\times 50\times 1~\mu$m$^3$ we would calculate an effective mass (see e.g.~\cite{Pinard1999,Gigan2006}) of $50\pm 5$~ng. Again, the actual flat-top mode shape of our device results in a decreased mean square displacement (by approx.\ 6$\%$) compared to the ideal doubly-clamped beam. Taking this into account yields a final effective mass of $53\pm 5$~ng, which agrees to within 10$\%$ with the experimentally observed value of $43\pm 2$~ng.

\begin{figure}[htbp]
\centerline{\includegraphics[width=0.48\textwidth]{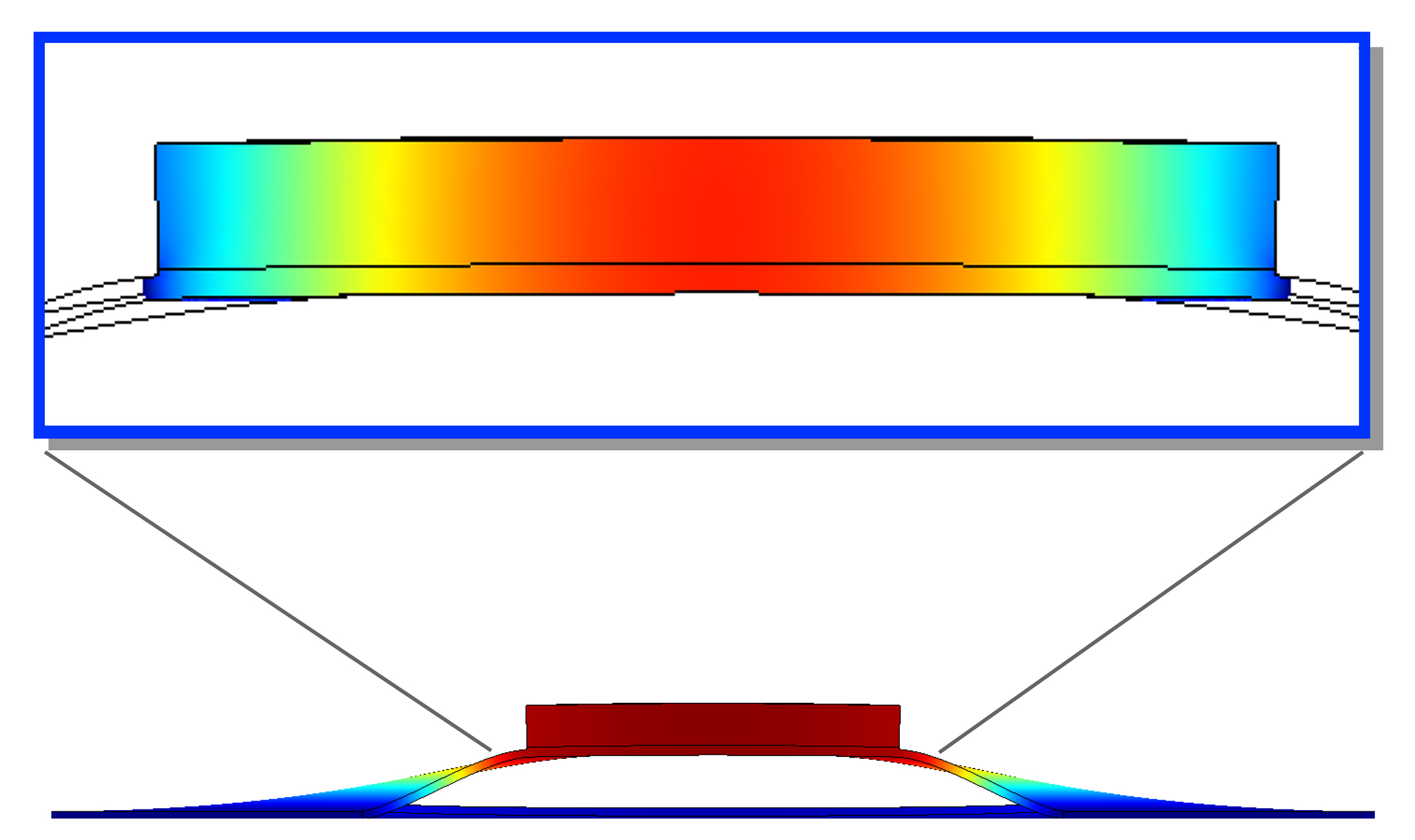}} \caption{FEM simulation of our optomechanical device. Shown is the side-view of the fundamental resonance mode at its maximum displacement (below). The cylindrical mirror pad on top of the Si$_3$N$_4$ beam induces a flat-top mode shape (inset).} \label{figs1}
\end{figure}

The abovementioned FEM simulations make use of the exact geometry and material data for our resonator. The main idea is to impose a force on the structure and have the FEM simulation calculate the deflection. Using Hooke's law one can then extract the spring constant $k$ of the device. The mode mass can be extracted by using $\omega_m=\sqrt{k/m_{mode}}$. For our specific device the FEM solver provides us with a spring constant of $2,196$~N/m and a fundamental mode at $\omega_m=2\pi\times 945$~kHz, which results in $m_{mode}=57\pm 5$~ng.\\

\begin{center}
\small\textbf{Error analysis}\\
\end{center}
\normalsize

The error associated with the noise power spectra peak areas, which provide the mechanical mean square displacement, can be estimated as follows:\ Assuming that the NPS comprises a sequence of $N$ independent data points ($x_i$, $y_i$) (with $i=1...N$) with measurement uncertainty ($\delta x_i$, $\delta y_i$) one can calculate the area underneath the NPS by Riemann integration as $A=\sum^{N-1}_{i=1}(x_{i+1}-x_i)y_i$ with an uncertainty $\delta A=\sqrt{\sum^{N-1}_{i=1}(x_{i+1}-x_i)^2(\delta y_i)^2}$, which is obtained by Gaussian error propagation and neglecting the uncertainty in $x$. The strongly cooled NPS shown in Figure 3a is given by a data set of $N=5,000$ points with $x_{i+1}-x_i=100$~Hz and with $\delta y_i\approx 1\times 10^{-34}$~m$^2$ Hz$^{-1}$ for all $i$. We obtain $A=3.780\times 10^{-28}$~m$^2$ (by numerically integrating the data set), $\delta A\approx\sqrt{N}\times 100$~Hz$\times 1\times 10^{-34}$~m$^2$ Hz$^{-1}=7.1\times 10^{-31}$~m$^2$ and an integrated noise floor of $N\times 100$~Hz$\times 7.3\times 10^{-34}$~m$^2$ Hz$^{-1}=3.65\times 10^{-28}$~m$^2$. This results in an integrated "real thermal noise" of $(3.78-3.65)\times 10^{-28}$~m$^2=1.3\times 10^{-29}$~m$^2$ with an overall error of approx.\ $\sqrt{2}\times 7.3\times 10^{-31}$~m$^2\approx 1\times 10^{-30}$~m$^2$, i.e.\ with an error of approx.\ $8\%$. The SNR of our measurement is therefore sufficient to support our result of $\langle n\rangle=32$ and accounts for an uncertainty of $\langle\delta n\rangle=\pm 1.5$.

Other possible sources of experimental uncertainty are:\ an uncertainty related to the absolute displacement amplitude calibration (amounting to approx.\ $12\%$ relative uncertainty), an uncertainty related to determining the mechanical resonance frequency (known up to an error of approx.\ $5\%$) and an uncertainty related to the absolute power calibration of the intracavity optical pump field (known up to an error of approx.\ $10\%$). These additional experimental uncertainties add up to an additional overall error of approx.\ $25\%$. All errors are conservatively estimated and finally result in $\langle n\rangle=32\pm 4$.\\

\begin{center}
\small\textbf{Shot-Noise}\\
\end{center}
\normalsize

The noise floor of our measurement is limited by optical shot-noise. The corresponding displacement noise can be calculated according to~\cite{BriantPhD} as
\begin{eqnarray}
\delta x_{Shot}=\frac{\lambda}{16F\sqrt{\frac{P\lambda}{hc}}}\cdot\sqrt{1+\left(\frac{\omega_m}{\kappa}\right)^2}\cdot\sqrt{\frac{T+l}{T}}\cdot\frac{P}{P_{MM}} \nonumber
\end{eqnarray}
Our experimental parameters (finesse $F=3,900$, input power $P=14~\mu$W, $\lambda=1064$~nm, $\omega_m=2\pi\times 945$~kHz, $\kappa=2\pi\times 770$~kHz, input coupler transmission $T=900$~ppm, overall intra-cavity losses $l=620$~ppm, optical input power (corrected for imperfect mode-matching) $P_{MM}=7~\mu$W) result in a minimal noise-floor of $\delta x_{Shot}=6\times 10^{-18}$~m Hz$^{-0.5}$.\\

\begin{acknowledgements}
We thank R. Lalezari (ATFilms) and M. Metzler, R. Ilic and M. Skvarla (CNF) and F. Blaser, T. Corbitt and W. Lang for discussion and support. We acknowledge support by the Austrian Science Fund FWF (Projects P19539, L426, START), by the European Commission (Projects MINOS, IQOS) and by the Foundational Questions Institute fqxi.org (Grants RFP2-08-03, RFP2-08-27). Part of this work was carried out at the Cornell NanoScale Facility, a member of the National Nanotechnology Infrastructure Network, which is supported by the National Science Foundation (Grant ECS-0335765). S.Gr. is a recipient of a DOC-fellowship of the Austrian Academy of Sciences and G.D.C. of a Marie Curie Fellowship of the European Commission. S.Gr. and M.R.V. are members of the FWF doctoral program Complex Quantum Systems (W1210).
\end{acknowledgements}

\end{document}